\documentclass[aps,reprint,prl,amsmath,floatfix,noeprint,superscriptaddress]{revtex4-2}

\usepackage{graphics}
\usepackage{natbib}
\usepackage{graphicx}
\usepackage{epsfig}
\usepackage{amsmath}
\usepackage{bm}
\usepackage{lineno}
\usepackage{color}
\usepackage{dcolumn}
\begin{document}
\title{Magnetoelastic Gilbert damping in magnetostrictive Fe$_{0.7}$Ga$_{0.3}$ thin films}
\author{W.~K.~Peria}
\affiliation{School of Physics and Astronomy, University of Minnesota, Minneapolis, Minnesota 55455, USA}
\author{X.~Wang}
\affiliation{Department of Materials Science and Engineering, University of Maryland, College Park, Maryland 20742, USA}
\author{H.~Yu}
\affiliation{Department of Materials Science and Engineering, University of Maryland, College Park, Maryland 20742, USA}
\author{S.~Lee}
\affiliation{Department of Materials Science and Engineering, University of Maryland, College Park, Maryland 20742, USA}
\affiliation{Department of Physics, Pukyong National University, Busan 48513, South Korea}
\author{I.~Takeuchi}
\affiliation{Department of Materials Science and Engineering, University of Maryland, College Park, Maryland 20742, USA}
\author{P.~A.~Crowell}
\email[Author to whom correspondence should be addressed: ]{crowell@umn.edu}
\affiliation{School of Physics and Astronomy, University of Minnesota, Minneapolis, Minnesota 55455, USA}
\begin{abstract}
We report an enhanced magnetoelastic contribution to the Gilbert damping in highly magnetostrictive Fe$_{0.7}$Ga$_{0.3}$ thin films. This effect is mitigated for perpendicular-to-plane fields, leading to a large anisotropy of the Gilbert damping in all of the films (up to a factor of 10 at room temperature). These claims are supported by broadband measurements of the ferromagnetic resonance linewidths over a range of temperatures (5 to 400~K), which serve to elucidate the effect of both the magnetostriction \textit{and} phonon relaxation on the magnetoelastic Gilbert damping.
\end{abstract}
\maketitle
Among the primary considerations in the design of spintronics devices is Gilbert damping. However, a full understanding of the mechanisms which cause damping of magnetization dynamics in ferromagnets remains elusive. Reports of anisotropy in the Gilbert damping have proven to be useful tools in the understanding of the underlying mechanisms involved \cite{Gilmore2010,Chen2018,Li2019}, but there is much that is yet unclear. Studies of the temperature dependence also promise to be a uniquely powerful tool for a complete physical understanding \cite{Kumar2017,Khodadadi2020}, however, there are few such reports in existence.

Recently, it has been shown that spins can be coherently coupled over large distances ($\sim$1~mm) using magnon-phonon coupling \cite{An2020,Ruckriegel2020,Casals2020}. It is also well known that magnetization dynamics can be excited elastically through this phenomenon \cite{Weiler2011}, but its effect on Gilbert damping has been largely confined to theoretical calculations \cite{Suhl1998,Rossi2005,Vittoria2010,Streib2018} and lacks clear experimental validation. Furthermore, most studies have focused on yttrium iron garnet (YIG), which is weakly magnetostrictive.

In this Letter, we observe a large and anisotropic magnetoelastic contribution to the Gilbert damping in highly magnetostrictive Fe$_{0.7}$Ga$_{0.3}$ films through broadband measurements of the ferromagnetic resonance (FMR) linewidths over a wide range of temperatures. The perpendicular-to-plane linewidths exhibit a relatively low minimum in the Gilbert damping of approximately 0.004, similar to that of bcc Fe \cite{Schoen2016}. At room temperature, the Gilbert damping is as large as a factor of 10 greater with field applied in plane relative to out of plane. In fact, for any given sample and temperature, the anisotropy is, at minimum, about a factor of 2. We argue this is due to a mitigation of the magnetoelastic contribution for perpendicular magnetization, arising from finite-thickness boundary conditions and weak elastic coupling to the substrate. The nonmonotonic temperature dependence of the Gilbert damping also shows the competing effects of the magnetostriction, which increases at low temperature, and the phonon viscosity, which generally decreases at low temperature.

The Fe$_{0.7}$Ga$_{0.3}$ films studied in this letter were deposited on SiO$_2$/Si wafers at room temperature by dc magnetron sputtering of an Fe$_{0.7}$Ga$_{0.3}$ target. The base pressure of the deposition chamber was $5 \times 10^{-8}$~torr, and the working pressure was kept at $5\times10^{-3}$ torr with Ar gas. The composition of the Fe$_{0.7}$Ga$_{0.3}$ films was quantitatively analyzed by energy dispersive spectroscopy (EDS). Films were grown with thicknesses of 21~nm, 33~nm, 57~nm, and 70~nm (the 21~nm, 57~nm, and 70~nm belong to the same growth series). An additional 33~nm film was grown at 200~$^\circ$C. The 33~nm room temperature deposition was etched using an ion mill to obtain films with thicknesses of 17~nm and 26~nm. The thicknesses of the films were measured using x-ray reflectometry (see Supplemental Material).

\begin{figure}
    \centering
    \includegraphics{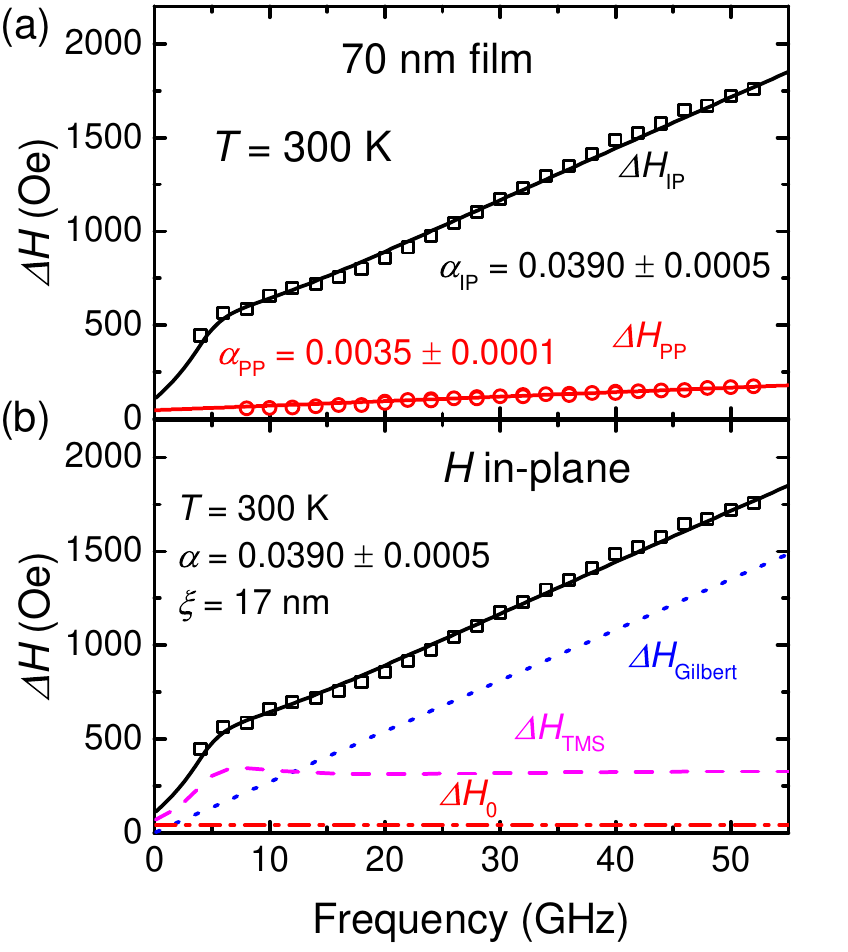}
    \caption{(a)~FMR linewidths for IP (black squares) and PP (red circles) configurations for the 70~nm film. The IP linewidths are fit to a model of two-magnon scattering and the PP linewidths are fit using the standard Gilbert damping model. (b)~Total linewidth (solid black), Gilbert linewidth (dotted blue), two-magnon scattering linewidth (dashed magenta), and inhomogeneous broadening (dashed/dotted red) for the 70~nm film with IP field.}
    \label{fig:70nmlinewidths}
\end{figure}

The FMR linewidths were measured using a setup involving a coplanar waveguide and modulation of the applied magnetic field for lock-in detection as described in Ref.\ \cite{Peria2020}. Measurements were done with the field applied in the plane (IP) and perpendicular to the plane (PP) of the film. The sample temperature was varied from 5~K to 400~K for both IP and PP configurations \footnote{Using a probe designed by NanOsc for a Quantum Design Physical Property Measurement System (PPMS).} with microwave excitation frequencies up to 52~GHz. The resonance fields and linewidths were isotropic in the plane, and the absence of in-plane magnetic anisotropy was verified with vibrating sample magnetometry (see Supplemental Material). This is also consistent with the abundance of grain boundaries observed with atomic force microscopy (AFM). In analyzing the FMR linewidths, we consider three contributions: Gilbert damping $4 \pi \alpha f / \gamma$ ($\alpha$ is the Gilbert damping coefficient, $f$ is the microwave frequency, and $\gamma$ is the gyromagnetic ratio), inhomogeneous broadening $\Delta H_0$, and two-magnon scattering $\Delta H_{TMS}$ (for IP fields). Eddy current damping and radiative damping contributions \cite{Schoen2015} are neglected because we expect them to be small ($< 10^{-4}$) for these films. Linewidths of the 70~nm film at 300~K for both configurations of the applied field are shown in Fig.\ \ref{fig:70nmlinewidths}(a), and the IP linewidths with individual contributions to the linewidth plotted separately in Fig.\ \ref{fig:70nmlinewidths}(b). We fit the IP linewidths using a model of two-magnon scattering based on granular defects \cite{McMichael2004,Krivosik2007,Peria2020}. The fit for the 70~nm film is shown in Fig.\ \ref{fig:70nmlinewidths}(b), along with the two-magnon contribution alone given by the magenta curve. The fit parameters are the Gilbert damping $\alpha$ (indicated on the figure) and the RMS inhomogeneity field $H'$. The defect correlation length $\xi$ is fixed to 17~nm based on the structural coherence length obtained with x-ray diffraction (XRD), which agrees well with the average grain diameter observed with AFM (see Supplemental Material). Furthermore, the high-frequency slope of the linewidths approaches that of the Gilbert damping since the two-magnon linewidth becomes constant at high frequencies [see Fig.\ \ref{fig:70nmlinewidths}(b)].

We now compare the IP and PP linewidths of the 70~nm film shown in Fig.\ \ref{fig:70nmlinewidths}(a). The two-magnon scattering mechanism is inactive with the magnetization perpendicular to the plane \cite{Arias1999}, and so the PP linewidths are fit linearly to extract the Gilbert damping. We obtain a value of $0.0035 \pm 0.0001$ for PP fields and $0.039 \pm 0.0005$ for IP fields, corresponding to an anisotropy larger than a factor of 10. \citet{Li2019} recently reported a large anisotropy ($\sim$ factor of 4) in epitaxial Co$_{50}$Fe$_{50}$ thin films.
\begin{figure}
  \includegraphics{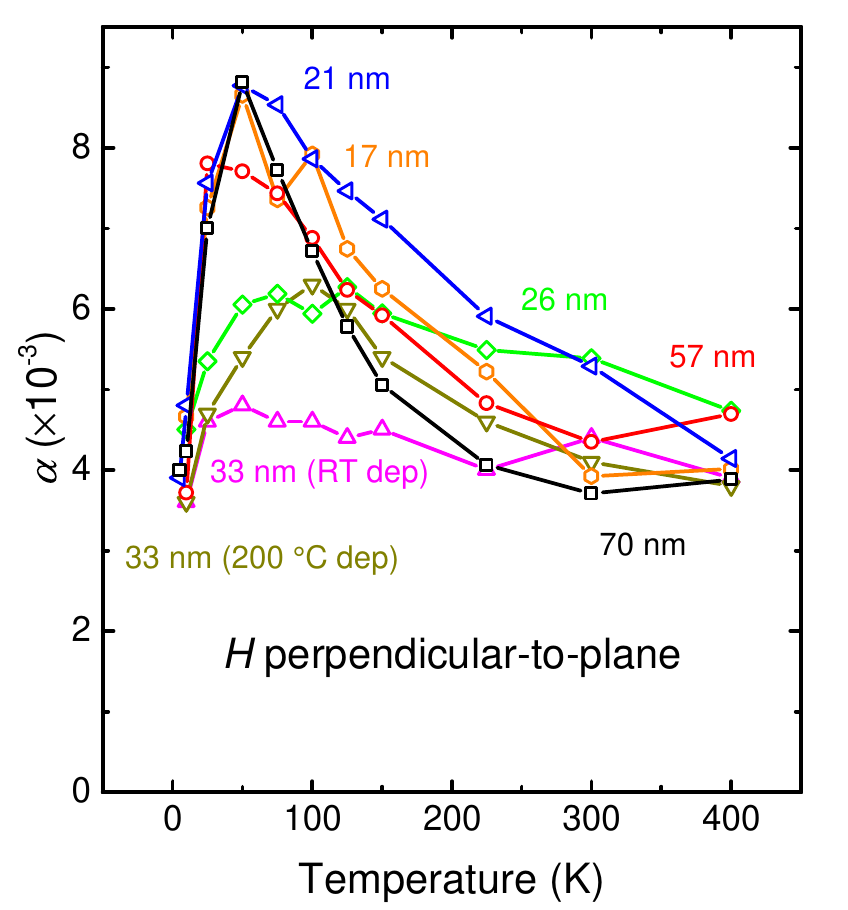}
  \caption{Gilbert damping $\alpha$ for PP field shown as a function of temperature for the 17~nm (orange), 21~nm (blue), 26~nm (green), 33~nm room temperature deposition (magenta), 33~nm 200~$^\circ$C deposition (gold), 57~nm (red), and 70~nm (black) Fe$_{0.7}$Ga$_{0.3}$ films.}\label{fig:PPalpha}
\end{figure}

First we discuss the dependence of the PP Gilbert damping $\alpha_{PP}$ on temperature for all of the films, shown in Fig.\ \ref{fig:PPalpha}. We observe a significant temperature dependence in all cases (with the exception of the 33~nm room temperature deposition), characterized by a maximum at around 50~K. Then, at the lowest temperatures (5 to 10~K), $\alpha_{PP}$ approaches the same value for all of the films ($\simeq0.004$).

\begin{figure}
  \centering
  \includegraphics{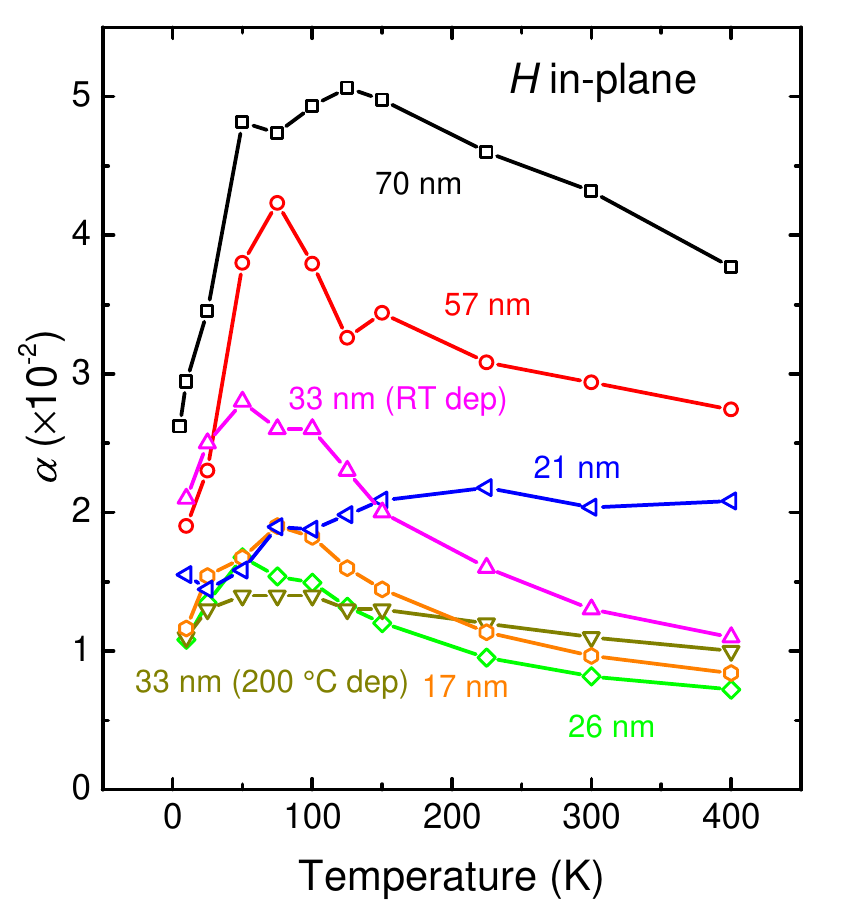}
  \caption{Gilbert damping $\alpha$ for IP field shown as a function of temperature for the 17~nm (orange), 21~nm (blue), 26~nm (green), 33~nm room temperature deposition (magenta), 33~nm 200~$^\circ$C deposition (gold), 57~nm (red), and 70~nm (black) Fe$_{0.7}$Ga$_{0.3}$ films.}\label{fig:IPalpha}
\end{figure}

Now we turn to the temperature dependence of the IP Gilbert damping $\alpha_{IP}$ shown in Fig.\ \ref{fig:IPalpha}. The values obtained here were obtained by fitting the linewidths linearly, but excluding the low-frequency points ($\lesssim 20$~GHz) since the two-magnon scattering becomes constant at high frequencies \cite{Woltersdorf2004}. Here we note, upon comparison with Fig.\ \ref{fig:PPalpha}, that a large anisotropy of the Gilbert damping exists for all of the samples. In the 70~nm film, for instance, $\alpha_{IP}$ is more than a factor of 10 larger than $\alpha_{PP}$ at 300~K. In the temperature dependence of $\alpha_{IP}$, we observe behavior which is similar to that seen in $\alpha_{PP}$ (Fig.\ \ref{fig:PPalpha}), namely, a maximum at around 50~K (with the exception of the 21~nm film). Here, however, $\alpha_{IP}$ does not approach a common value at the lowest temperatures in all of the samples as it does in the PP case.

The IP Gilbert damping is larger than the PP Gilbert damping for all of the samples over the entire range of temperatures measured. This anisotropy of the Gilbert damping---along with the nonmonotonic temperature dependence---in all seven samples implies a contribution to the Gilbert damping in addition to Kambersk\'y damping. We have verified that the orientation of FeGa(110) planes is completely random with XRD for the 33~nm (both depositions) and 70~nm films (see Supplemental Material), and it is therefore not possible that the anisotropy is due to Kambersk\'y damping. Interface anisotropy has reportedly led to anisotropic Kambersk\'y damping in ultrathin ($\sim$1~nm) films of Fe \cite{Chen2018}, but this is highly unlikely in our case due to the relatively large thicknesses of the films. In addition, the fact that the damping anisotropy shows no clear correlation with film thickness furthers the case that intrinsic effects, which tend to show a larger anisotropy in thinner films \cite{Chen2018}, cannot be the cause. The longitudinal resistivity $\rho_{xx}$ of the 33~nm (both depositions) and 70~nm films (see Supplemental Material) shows very weak temperature dependence. In the Kambersk\'y model, the temperature dependence of the damping is primarily determined by the electron momentum relaxation time $\tau$, and we would therefore not expect the Kambersk\'y damping to show a significant temperature dependence for samples where the residual resistivity ratio is approximately unity. It is plausible that the Kambersk\'y damping would still show a temperature dependence in situations where the spin polarization is a strong function of temperature, due to changes in the amount of interband spin-flip scattering. This kind of damping, however, would be expected to decrease at low temperature \cite{Kambersky2007,Gilmore2007}. The temperature dependence we observe for both $\alpha_{PP}$ and $\alpha_{IP}$ is therefore inconsistent with Kambersk\'y's model, and the similarity between the two cases in this regard suggests that the enhanced Gilbert damping has a common cause that is mitigated in the PP configuration.

It has been proposed that magnetoelastic coupling can lead to Gilbertlike magnetization damping through phonon relaxation processes \cite{Kittel1958,Suhl1998,Vittoria2010}. Similar treatments calculate the magnetoelastic energy loss through interaction with the thermal population of phonons \cite{Rossi2005,Ruckriegel2014}.
The Kambersk\'y mechanism is often assumed to be the dominant Gilbert damping mechanism in metallic samples, so magnetoelastic Gilbert damping is usually studied in magnetic insulators, particularly yttrium iron garnet (YIG). There is the possibility, however, for the magnetoelastic damping to dominate in metallic samples where the magnetostriction is large, such as in Fe-Ga alloys. Later we will discuss how magnetoelastic damping can be mitigated in thin films by orienting the magnetization perpendicular to the plane, and how the degree to which it is mitigated depends on the boundary conditions of the film.

\begin{figure}
  \centering
  \includegraphics{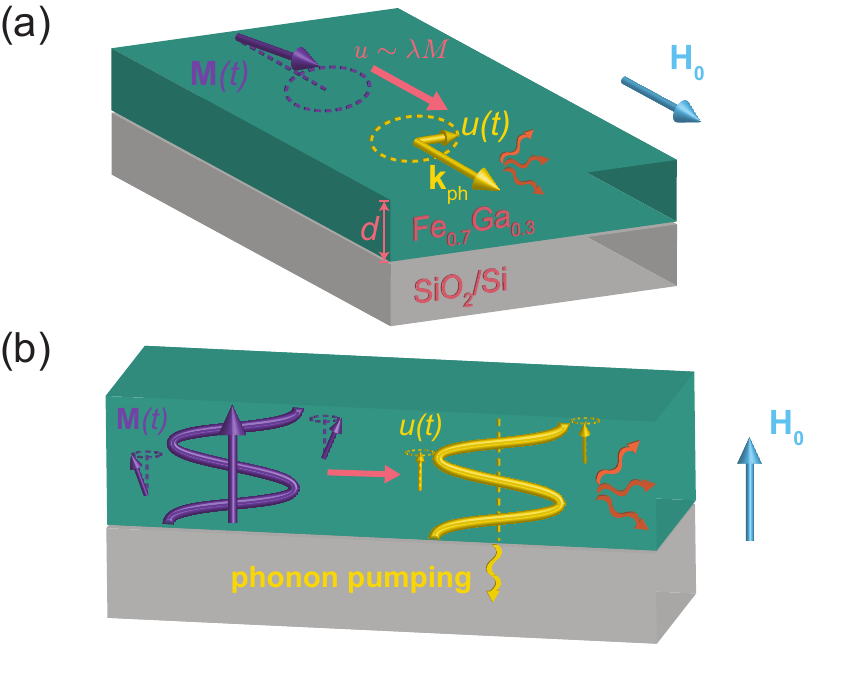}
  \caption{(a) Depiction of magnetoelastic damping process for magnetization in plane and (b) perpendicular to plane, where $\mathbf{M}(t)$ is the magnetization vector and $u(t)$ is the lattice displacement. In panel (b), the magnon-phonon conversion process is suppressed when $d < \pi / k_{ph}$, where $d$ is the film thickness and $k_{ph}$ is the transverse phonon wavenumber at the FMR frequency.}\label{fig:theory}
\end{figure}

Here we outline a theory of magnetoelastic damping, which relies on the damping of magnetoelastic modes through phonon relaxation mechanisms. Figure \ref{fig:theory} illustrates the flow of energy through such a process. Analytically, the procedure is to equate the steady-state heating rate due to Gilbert damping to the heating rate due to crystal viscosity, and solve for the Gilbert damping $\alpha$ in terms of the crystal shear viscosity $\eta$ and the magnetostrictive coefficients $\lambda_{hkl}$. Shear strain $u_{ij}$ resulting from the magnetoelastic interaction can be expressed as $u_{ij} = \lambda_{111} m_i m_j$ \cite{LandauEM}, where $m_i \equiv M_i / M_s$ are the reduced magnetizations. The leading-order shears thus have equations of motion given by $\dot{u}_{iz} = \lambda_{111} \dot{m}_i$, where $i = x$ or $y$, and $z$ is the direction of the static magnetization so that $m_z \approx 1$. Longitudinal modes are quadratic in the dynamical component of the magnetization \cite{Kittel1958} and so will be neglected in this analysis.

The heating rate due to Gilbert damping can be written as $\dot{Q}_\alpha = \frac{M_s}{\gamma} \alpha (\dot{m}_x^2+\dot{m}_y^2)$, and the heating rate due to the damping of phonon modes as $\dot{Q}_\eta = 4 \eta  ( \dot{u}_{x z}^2 + \dot{u}_{y z}^2 ) = 4 \eta \lambda_{111}^2 (\dot{m}_x^2+\dot{m}_y^2)$ \cite{Vittoria2010}, with the factor of 4 accounting for the symmetry of the strain tensor. Equating the two, and solving for $\alpha$ (henceforward referred to as $\alpha_{me}$), we obtain
\begin{equation}\label{eq:alphame}
  \alpha_{me} = \frac{4 \gamma}{M_s} \eta \lambda_{111}^2~.
\end{equation}
We will restrict our attention to the case of isotropic magnetostriction, and set $\lambda_{111} = \lambda$.

In order to use Eq.\ \ref{eq:alphame} to estimate $\alpha_{me}$ in our films, we first estimate the shear viscosity, given for transverse phonons with frequency $\omega$ and relaxation time $\tau$ as \cite{LandauToe}
\begin{equation}\label{eq:viscosity}
  \eta = \frac{2 \rho c_t^2}{\omega^2 \tau}~,
\end{equation}
where $\rho$ is the mass density and $c_t$ is the transverse speed of sound. Using $\omega/2\pi = 10$~GHz, $\tau = 10^{-11}$~s, and $c_t = 2.5$~km/s, we obtain $\eta \approx 2.3$~Pa s. (The estimate of the phonon relaxation time is based on a phonon mean free path of the order of the grain size: $\sim$10~nm.) Furthermore, the magnetostriction of an equivalent sample has been measured to be $\sim$100~ppm at room temperature \cite{Hattrick-Simpers2008}. Then, with $\gamma / 2 \pi = 29$~GHz/T and $M_s = 1123$~emu/cc (extracted from FMR data taken at 300~K), we estimate $\alpha_{me} \approx 0.016$. This estimate gives us immediate cause to suspect that magnetoelastic Gilbert damping is significant (or even dominant) in these films.

We now discuss why the magnetoelastic damping can be much weaker for PP magnetization in sufficiently thin films. We will start by assuming that there is no coupling between the film and substrate, and later we will relax this assumption. In this case the only phonons excited by the magnetization, to leading-order in the magnetizations and strains, are transverse modes propagating in the direction of the static magnetization \cite{Kittel1958}. One may assume that the minimum allowable phonon wavenumber is given by $\pi / d$, where $d$ is the film thickness, since this corresponds to the minimum wavenumber for a substrate having much lower acoustic impedance than the film (requiring the phonons to have antinodes at the interfaces) \cite{Streib2018}. (We also assume an easy-axis magnetic anisotropy at the interfaces, so that the dynamical magnetizations have antinodes at the interfaces.) We expect then that the magnetoelastic damping will be suppressed for cases where the phonon wavelength, at the frequency of the precessing magnetization, is greater than twice the film thickness [see Fig.\ \ref{fig:theory}(b)]. Thus, in sufficiently thin films (with weakly-coupled substrates), the magnetoelastic damping process can be suppressed when the magnetization is perpendicular to the plane. However, the magnetoelastic damping can be active (albeit mitigated) when there is nonnegligible or ``intermediate'' coupling to the substrate.

Before moving on, we briefly note the implications of Eq.\ (\ref{eq:alphame}) for the temperature dependence of the Gilbert damping. On the basis of the magnetostriction alone, $\alpha_{me}$ would be expected to increase monotonically as temperature is decreased ($\lambda$ has been shown to increase by nearly a factor of 2 from room temperature to 4~K in bulk samples with similar compositions \cite{Clark2005}). However, the viscosity $\eta$ would be expected to decrease at low temperature, leading to the possibility of a local maximum in $\alpha_{me}$. In polycrystalline samples where the grain size is smaller than the phonon wavelength, viscous damping of phonons due to thermal conduction caused by stress inhomogeneities can be significant \cite{Zener1938,LandauToe}. (In our case the phonon wavelengths are $\sim100$~nm and the grain sizes are $\sim10$~nm.) This effect scales with temperature as $\eta \sim T \alpha_T^2 / C \chi$ \cite{Zener1938}, where $\alpha_T$ is the thermal expansion coefficient, $C$ is the specific heat at constant volume, and $\chi$ is the compressibility. At higher temperatures, $\alpha_T$ and $C$ will approach constant values, and $\chi$ will always depend weakly on temperature. We therefore expect that the viscosity is approximately linear in $T$. In this case, $\alpha_{me}$ is maximized where $\lambda^2(T)$ has an inflection point.

We proceed to explain our data in terms of the mechanism described above, turning our attention again to the PP Gilbert damping for all of the films shown in Fig.\ \ref{fig:PPalpha}.
\begin{figure}
  \centering
  \includegraphics{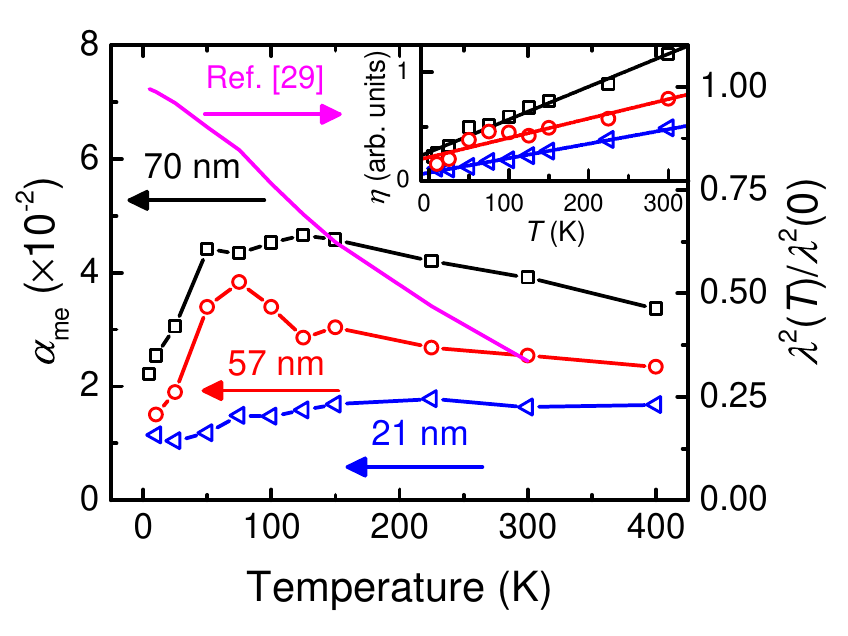}
  \caption{Magnetoelastic Gilbert damping $\alpha_{me}$ for the 21~nm (blue), 57~nm (red), and 70~nm (black) films (left ordinate) and $\lambda^2(T)/\lambda^2(0)$ from \citet{Clark2005} (magenta; right ordinate) shown as a function of temperature. Inset shows the ratio of $\alpha_{me}$ and $\lambda^2(T)/\lambda^2(0)$, labeled as $\eta(T)$, along with linear fits for the 21~nm (blue), 57~nm (red), and 70~nm (black) films.}\label{fig:correlation}
\end{figure}
 We previously argued that the magnetoelastic damping mechanism will be suppressed for the case where the acoustic impedances of the film and substrate are mismatched. However, the clear dependence on temperature, which we have already shown is inconsistent with Kambersk\'y damping, appears to be consistent with the magnetoelastic damping mechanism. We estimate that the acoustic impedance of the film (defined as the product of mass density $\rho$ and transverse speed of sound $c_t$ \cite{Streib2018}) is about a factor of 2 larger than the substrate. This suggests that the elastic coupling between the film and substrate, albeit weak, may be nonnegligible. Furthermore, experiments with YIG/GGG heterostructures (where the acoustic match is good) have demonstrated magnetic excitation of phononic standing waves that have boundary conditions dictated by the combined thickness of the film and substrate, rather than the film thickness alone (i.e., the wavelengths are much larger than the film thickness) \cite{Ye1988,An2020}. In this case, the Gilbert damping may contain some contribution from the magnetoelastic mechanism. A final point is that $\alpha_{PP}$ approaches $\simeq 0.004$ at 5 to 10~K for all of the films. Both the magnetostriction and the viscosity are quantities which could have significant variation between samples, leading to variations in $\alpha_{me}$. However, the viscosity becomes small at low temperature, which means that the Gilbert damping will approach the Kambersk\'y ``limit,'' a property that is determined by the electronic structure, implying that the Kambersk\'y damping is $\simeq 0.004$ in these films and that it is the primary contribution to the Gilbert damping near $T = 0$.

Now we revisit the IP Gilbert damping shown in Fig.\ \ref{fig:IPalpha}. In this configuration, there is a strong temperature dependence of the Gilbert damping similar to that of the PP case, again implying the presence of magnetoelastic damping. However, the overall magnitude is much higher. That is because in this case arbitrarily long wavelength phonons can be excited regardless of the thickness of the film. Although we cannot directly measure the magnetostriction as a function of temperature, we estimate the scaling behavior of $\lambda$ by interpolating the data in Ref.\ \cite{Clark2005} taken for bulk samples of similar composition. In order to demonstrate that $\alpha_{IP}$ scales with temperature as expected from the model, we have plotted the quantities $\alpha_{me}$ and $\lambda^2(T)/\lambda^2(0)$ as functions of temperature in Fig.\ \ref{fig:correlation}---where we define the quantity $\alpha_{me} \equiv \alpha_{IP} - 0.004$---for the 21~nm, 57~nm, and 70~nm films (which are part of the same growth). The correlation between the two quantities is not completely convincing. There is, however, an additional temperature dependence in $\alpha_{me}$ besides $\lambda^2(T)$, namely, the viscosity $\eta (T)$. The inset of Fig.\ \ref{fig:correlation} shows the ratio of $\alpha_{me}$ and $\lambda^2(T)$, which [from Eq.\ (\ref{eq:alphame})] is proportional to $\eta (T)$. The linear fits provide strong evidence that the mechanism behind the viscosity is indeed the thermal conduction process that we have argued is approximately linear in $T$. It is noteworthy that the maximum in $\alpha_{me}$ ($\sim 50$ to $75$~K for all of the samples) coincides approximately with the inflection point in $\lambda^2(T)$. This was a consquence of our assumption that $\eta(T)$ should be roughly linear. We also obtain a significant value for the zero-temperature viscosity, which is around 25~\% of the value at 300~K. This is likely due to boundary-scattering processes which will prevent $\alpha_{me}$ from going to zero at low temperatures, particularly for in-plane magnetization where $\alpha_{me}$ is much larger than 0.004 (our estimate for the Kambersk\'y damping). For the PP case, $\alpha_{me}$ is much smaller due to limitations on the wavelengths of phonons that can be excited, so the Gilbert damping of all the samples approaches the Kambersk\'y limit of 0.004 near zero temperature. We also found that $\eta (T)$ was linear for the 33~nm (200~$^\circ$C deposition) film, but had a more complicated dependence on $T$ for the 17~nm, 26~nm, and 33~nm (room temperature deposition) films (the latter three being notably of the same growth). The viscosity near zero temperature is within roughly a factor of 2 for all seven of the samples, however.

Finally, we propose that this mechanism may be responsible for a Gilbert damping anisotropy of similar magnitude reported in Ref.\ \cite{Li2019}, observed in an epitaxial Co$_{0.5}$Fe$_{0.5}$ thin film. The authors attributed the anisotropy to the Kambersk\'y mechanism \cite{Kambersky1970,Kambersky1976,Kambersky2007,Gilmore2007}, arising from tetragonal distortions of the lattice. The magnetostriction is known to be highly anisotropic in bulk Co$_{0.5}$Fe$_{0.5}$, \textit{viz.}, $\lambda_{100} = 150$~ppm and $\lambda_{111} = 30$~ppm \cite{Hall1960}. We therefore expect that the Gilbert damping arising from the mechanism we have described may be much larger for $\mathbf{M} \parallel (110)$ than $\mathbf{M} \parallel (100)$, which is precisely what the authors observed.

In summary, we observe large and anisotropic magnetoelastic Gilbert damping in Fe$_{0.7}$Ga$_{0.3}$ polycrystalline thin films (thicknesses ranging from 17 to 70 nm). At 300~K, the damping coefficient is more than a factor of 10 larger for field in plane than it is for field perpendicular to the plane in the 70~nm film. The large anisotropy is caused by a mitigation of the magnetoelastic effect for perpendicular-to-plane fields due to a dependence on the elastic coupling of the film to the substrate, which in our case is weak. Finally, there is a nonmonotonic temperature dependence of the Gilbert damping, which we show is consistent with our model.
\begin{acknowledgments}
We acknowledge Rohit Pant and Dyland Kirsch for assistance with thin film deposition and characterization. This work was supported by SMART, a center funded by nCORE, a Semiconductor Research Corporation program sponsored by NIST. Parts of this work were carried out in the Characterization Facility, University of Minnesota, which receives partial support from NSF through the MRSEC program, and the Minnesota Nano Center, which is supported by NSF through the National Nano Coordinated Infrastructure Network, Award Number NNCI~-~1542202.
\end{acknowledgments}
%
%
%
\end{document}


\title{Supplemental Material for\\ ``Magnetoelastic Gilbert damping in magnetostrictive Fe$_{0.7}$Ga$_{0.3}$ thin films''}
\author{W.~K.~Peria}
\affiliation{School of Physics and Astronomy, University of Minnesota, Minneapolis, Minnesota 55455, USA}
\author{X.~Wang}
\affiliation{Department of Materials Science and Engineering, University of Maryland, College Park, Maryland 20742, USA}
\author{H.~Yu}
\affiliation{Department of Materials Science and Engineering, University of Maryland, College Park, Maryland 20742, USA}
\author{S.~Lee}
\affiliation{Department of Materials Science and Engineering, University of Maryland, College Park, Maryland 20742, USA}
\affiliation{Department of Physics, Pukyong National University, Busan 48513, South Korea}
\author{I.~Takeuchi}
\affiliation{Department of Materials Science and Engineering, University of Maryland, College Park, Maryland 20742, USA}
\author{P.~A.~Crowell}
\affiliation{School of Physics and Astronomy, University of Minnesota, Minneapolis, Minnesota 55455, USA}
\maketitle
\tableofcontents
\section{Magnetization dynamics}\label{sec:dynamics}
The treatment of magnetization dynamics begins with the Landau-Lifshitz-Gilbert equation of motion
\begin{equation}
\frac{d \mathbf{M}}{dt} = - \gamma \mathbf{M} \times \mathbf{H}_{eff} + \frac{\alpha}{M_s} \mathbf{M} \times \frac{d \mathbf{M}}{dt}
\end{equation}
where the relaxation is characterized by the Gilbert damping parameter $\alpha$. Upon linearizing this equation in the dynamic component of the magnetization, one obtains for the ac magnetic susceptibility of the uniform $\mathbf{q}=0$ mode
\begin{equation}\label{eq:chi}
  \chi_{ac} (\mathbf{q} = 0, \omega) \propto \frac{\alpha \omega / \gamma}{(H - H_{FMR})^2 + (\alpha \omega / \gamma)^2}
\end{equation}
so that the field-swept full-width-at-half-maximum linewidth is given by $\Delta H_{FWHM} = 2 \alpha \omega / \gamma$. Therefore, the Gilbert damping parameter $\alpha$ is obtained by measuring $\Delta H_{FWHM}$ as a function of $\omega$.

Relaxation of the uniform mode can include mechanisms which are not described by Gilbert damping. The most common of these is inhomogeneous broadening, which results from inhomogeneities in the system and is constant as a function of frequency. Another mechanism is two-magnon scattering, which is also extrinsic in nature. Two-magnon scattering originates from the negative group velocity at low $\mathbf{q}$ of the backward volume mode magnons for in-plane magnetization. The negative group velocity is due to a lowering of the magnetostatic surface charge energy for increasing $\mathbf{q}$. The existence of negative group velocity at low $\mathbf{q}$ leads to the appearance of a mode at nonzero $\mathbf{q}$ that is degenerate with the uniform mode. Two-magnon scattering refers to the scattering of the uniform mode to the nonuniform degenerate mode.

Much work has been done on the treatment of two-magnon scattering \cite{Arias1999,McMichael2004,Krivosik2007}, and here we will simply give an expression for the contribution of two-magnon scattering to the field-swept linewidth
\begin{equation}\label{eq:TMSlinewidth}
  \Delta H_{TMS} = \frac{\gamma^2 \xi^2 H'^{2}}{d \omega / d H} \int d^2 \mathbf{q} ~ \Lambda_{0 \mathbf{q}} \frac{1}{(1 + (q \xi)^2)^{3/2}} \frac{1}{\pi} \frac{\omega \alpha}{(\omega \alpha)^2 + (\omega - \omega_{FMR})^2}
\end{equation}
with $\xi$ the defect correlation length, $H'$ the RMS inhomogeneity field, and $\Lambda_{0 \mathbf{q}}$ the magnon-magnon coupling. In general, this leads to a nonlinear dependence of the linewidth on frequency. Eq.\ \ref{eq:TMSlinewidth} is used to fit the IP linewidths.
\section{Ferromagnetic resonance linewidths of 70~\lowercase{nm} film}
\begin{figure}
  \centering
  \includegraphics{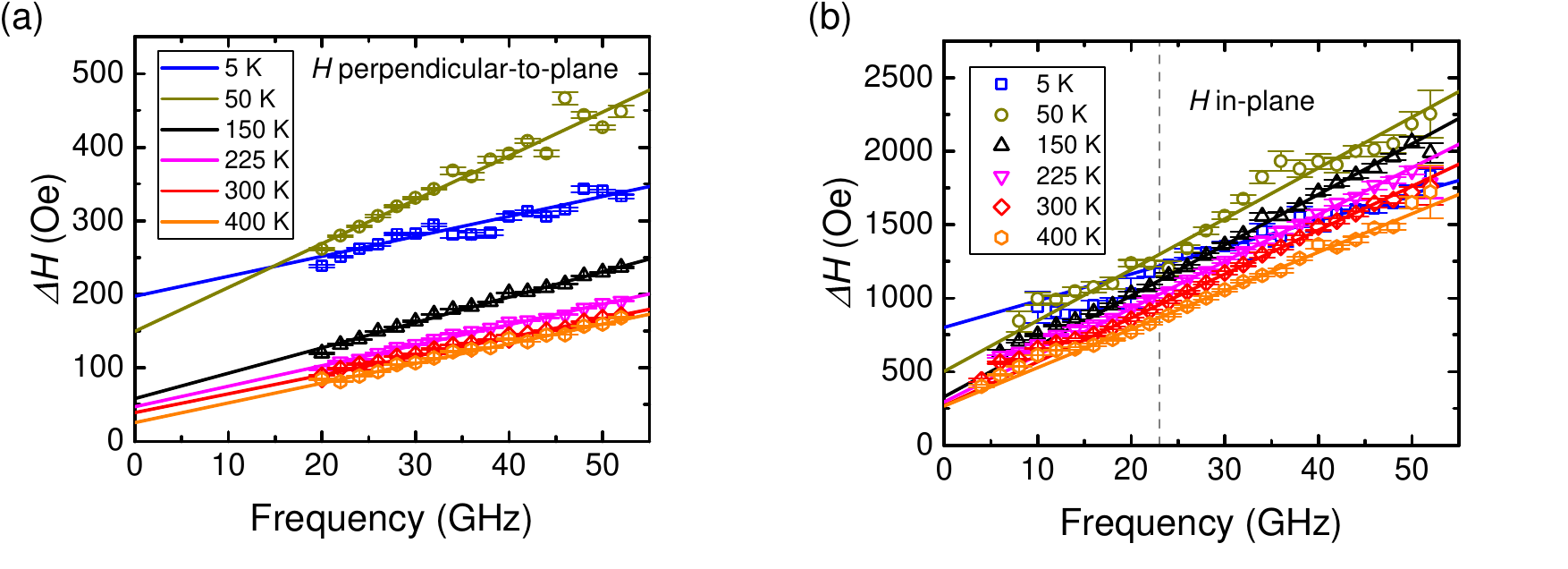}
  \caption{FMR linewidths of the 70~nm film with field PP (a) and IP (b) for sample temperatures of 5~K (blue), 50~K (gold), 150~K (black), 225~K (magenta), 300~K (red), and 400~K (orange). The solid lines are linear fits in both panels. In (b), the vertical dashed line indicates the lower bound of the points included in the fit.}\label{fig:70nmlinewidths}
\end{figure}

The field-swept FMR linewidths of the 70~nm film are shown in Fig.\ \ref{fig:70nmlinewidths} for field PP and IP. For the case of field IP, the data above 23~GHz were fit linearly to obtain the Gilbert damping. (This value varied between different samples since the characteristic roll-off frequency depends on both defect lengthscale and film thickness, but remained in the range 20 to 25~GHz.) It is safe to do this provided there are no inhomogeneities at lengthscales smaller than a few nm, which could cause the two-magnon scattering contribution to the linewidth to roll off at higher frequencies. We believe that defects at such small lengthscales are highly unlikely given the characterization performed on these samples.
\section{X-ray reflectivity}
\begin{figure*}
  \centering
  \includegraphics{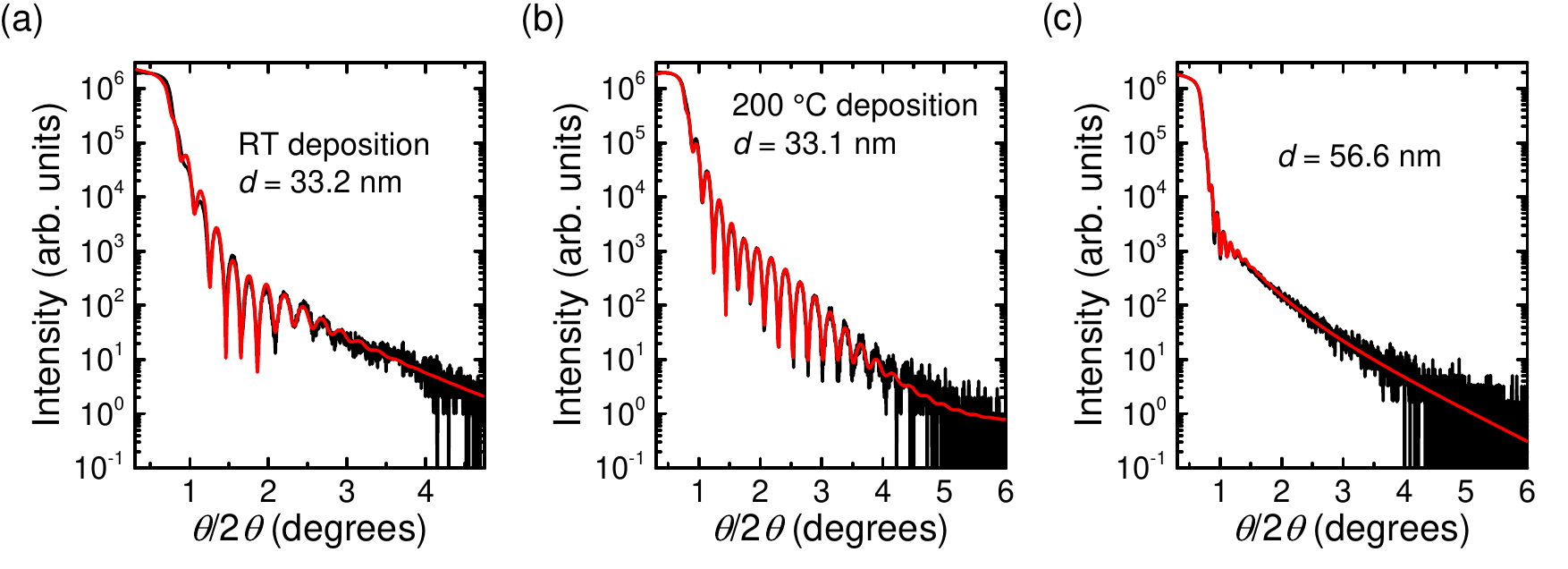}
  \caption{X-ray reflectivity data (black) overlaid with fits (red) for the (a) 33~nm (room temperature deposition), (b) 33~nm (200~$^\circ$C deposition), and (c) 57~nm films. Thicknesses $d$ obtained from the fits are indicated on the figure.}\label{fig:reflectivity}
\end{figure*}

In Fig.\ \ref{fig:reflectivity} we show x-ray reflectivity measurements at grazing incidence for 33~nm (room temperature and 200~$^\circ$C depositions) and 57~nm films. The measurements were taken using a Rigaku SmartLab diffractometer. The thicknesses $d$ yielded by the fits of the data are indicated on the figure.
\section{X-ray diffraction}
X-ray diffraction (XRD) measurements were performed in order to determine both the degree of orientation and the structural coherence length of the films.

\begin{figure*}
  \centering
  \includegraphics{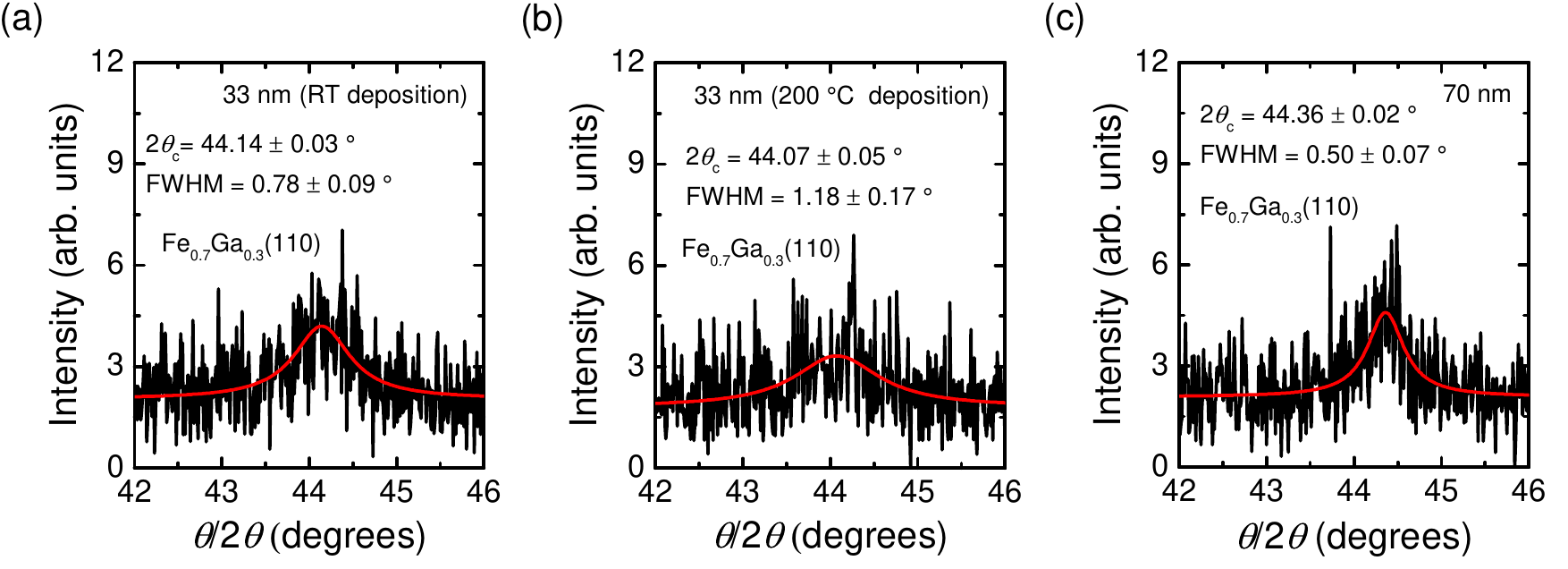}
  \caption{X-ray diffraction symmetric $\theta / 2 \theta$ scans for (a) 33 nm room temperature deposition, (b) 33 nm 200~$^\circ$C deposition, and (c) 70~nm films. Full width at half maxima (FWHM) and $2 \theta$ center positions are indicated on the figure.}\label{fig:coupled}
\end{figure*}

Symmetric $\theta/2\theta$ scans were taken with a Rigaku Smartlab diffractometer using Cu $K \alpha_1$ ($\lambda = 1.54$~\AA) radiation. The data for both samples are shown in Fig.\ \ref{fig:coupled}. The grain size was estimated using the Scherrer formula for spherical grains \cite{Birkholz2006} as 13~nm, 9~nm, and 17~nm for the 33~nm (room temperature deposition), 33~nm (200~$^\circ$C deposition), and 70~nm films respectively.

\begin{figure}
  \centering
  \includegraphics{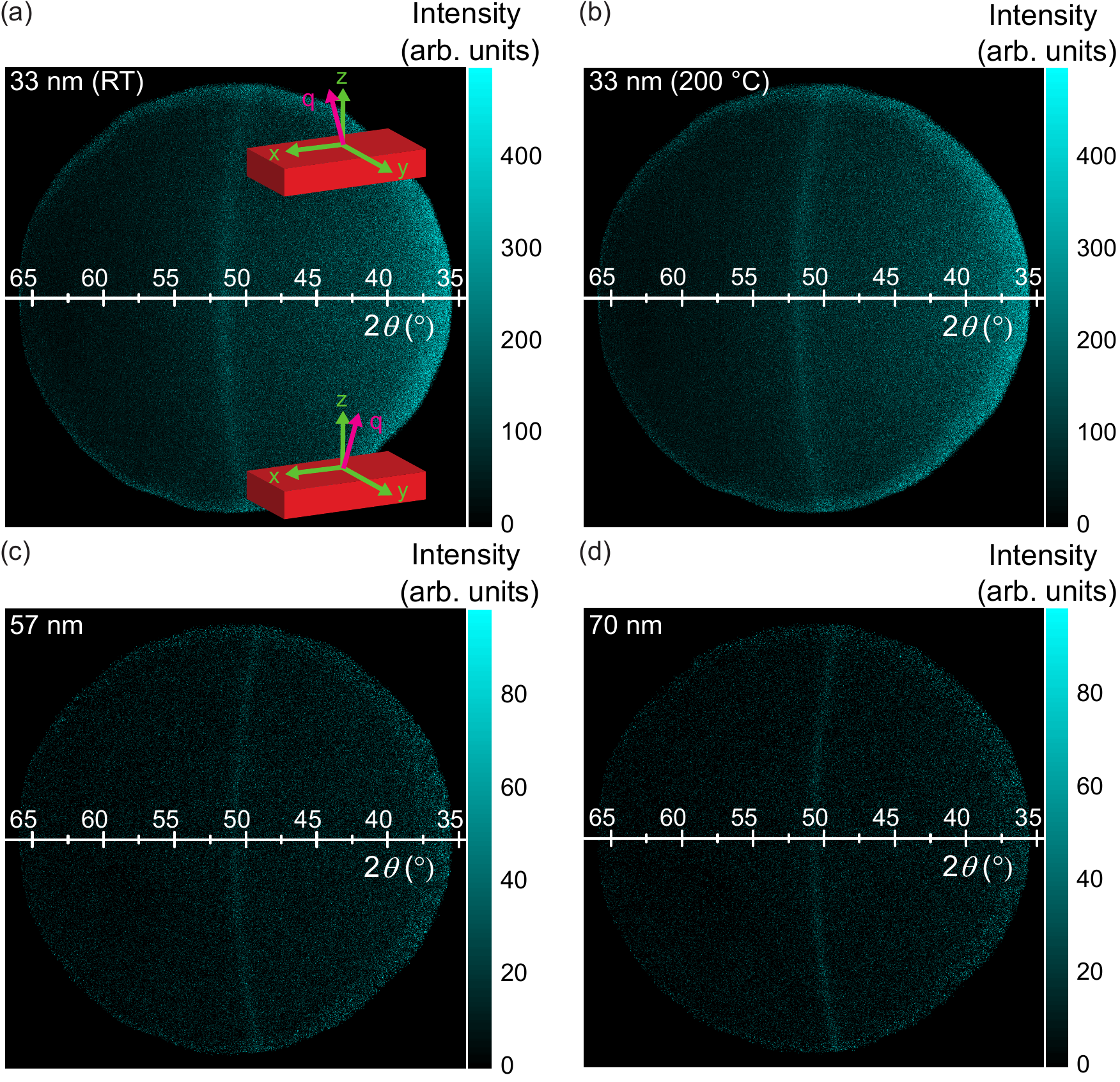}
  \caption{Two-dimensional detector images of the Fe$_{0.7}$Ga$_{0.3}$(110) peak for (a) 33~nm (room temperature deposition), (b) 33~nm (200 $^\circ$C deposition), (c) 57~nm, and (d) 70 nm films. The total scattering angle is 2$\theta$ and is shown on the abscissa. The measurement is conducted such that the symmetric configuration corresponds to the center of the detector, which is to say that the incident radiation is at an angle $\omega \simeq 26^\circ$ relative to the sample surface. In panel (a), the effect of moving vertically from the center of the detector on the scattering vector $\mathbf{q}$ is shown ($\mathbf{q}$ is canted into the $y$-$z$ plane).}\label{fig:2DXRD}
\end{figure}

Two-dimensional images were collected with a Bruker D8 Discover diffractometer using Co $K \alpha_1$ ($\lambda = 1.79$~\AA) radiation. Detector images showing the ``ring'' corresponding to the Fe$_{0.7}$Ga$_{0.3}$(110) peak in four different samples are shown in Fig.\ \ref{fig:2DXRD}. The ring indicates that the Fe$_{0.7}$Ga$_{0.3}$(110) planes are randomly oriented over the range of the detector, which we take to be evidence that there is no texture over a macroscopic scale in these samples. Furthermore, the films were grown directly on top of amorphous SiO$_2$ layers, so we do not expect an epitaxial relationship between the film and substrate. The Fe$_{0.7}$Ga$_{0.3}$(110) peaks were the only measurable Bragg peaks since the structure factor is highest for this case.
\section{Atomic force microscopy}
\begin{figure*}
  \centering
  \includegraphics{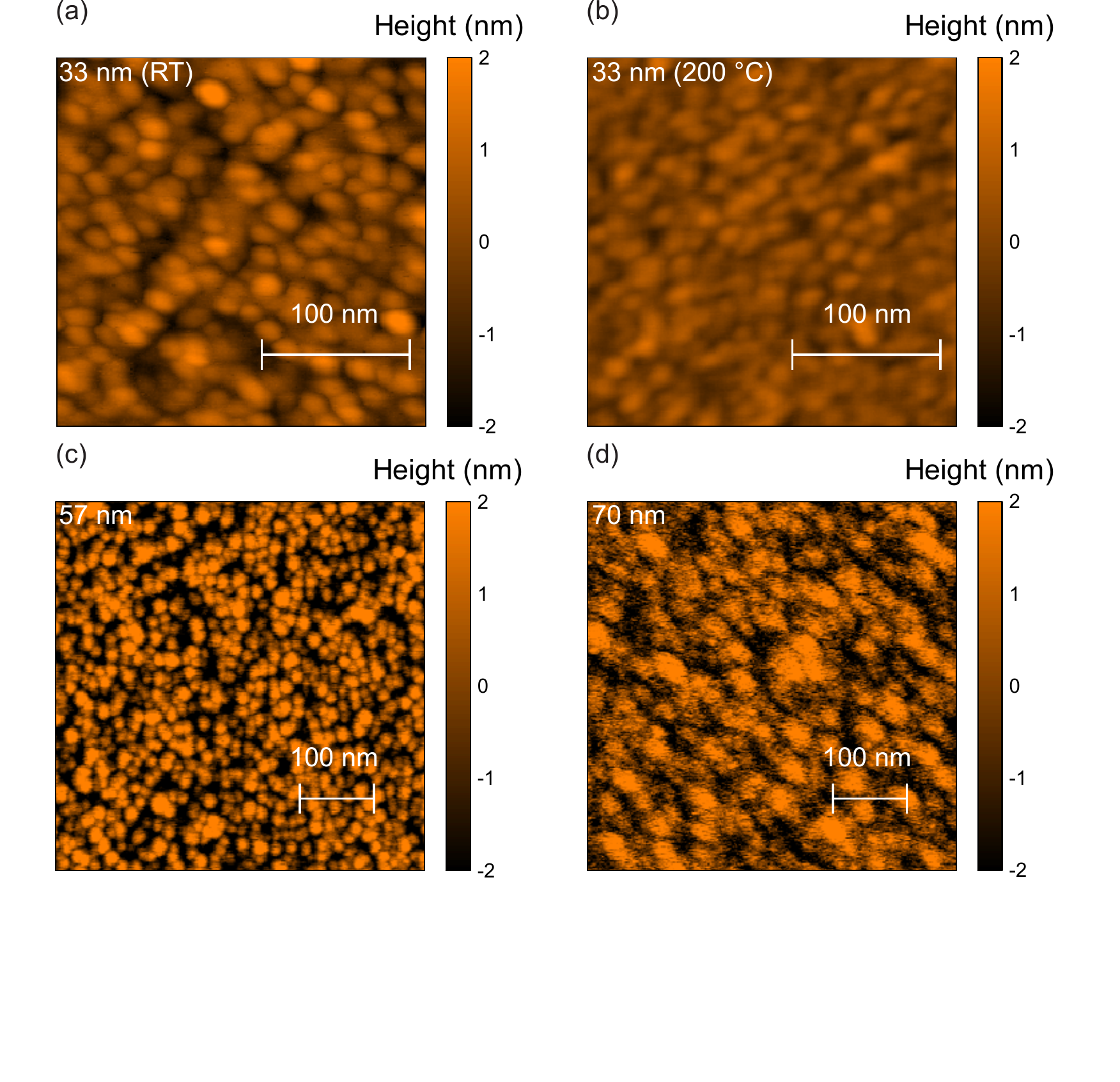}
  \caption{Atomic force microscopy for (a) 33~nm (room temperature deposition), (b) 33~nm (200 $^\circ$C deposition), (c) 57~nm, and (d) 70 nm films. RMS roughnesses are (a) 0.7~nm, (b) 0.4~nm, (c) 1.5~nm, and (d) 1.3~nm.}\label{fig:AFM}
\end{figure*}

Atomic force microscopy data are shown in Fig.\ \ref{fig:AFM} for the 33~nm (room temperature and 200~$^\circ$C depositions), 57~nm, and 70~nm films. The field-of-view is 250~nm for the 33~nm films and 500~nm for the 57~nm and 70~nm films. The root-mean-square (RMS) roughness of the sample surfaces is 0.7~nm, 0.4~nm, 1.5~nm, and 1.3~nm for the 33~nm (room temperature deposition), 33~nm (200~$^\circ$C deposition), 57~nm, and 70~nm films, respectively .
\section{Vibrating sample magnetometry}
\begin{figure}
  \centering
  \includegraphics{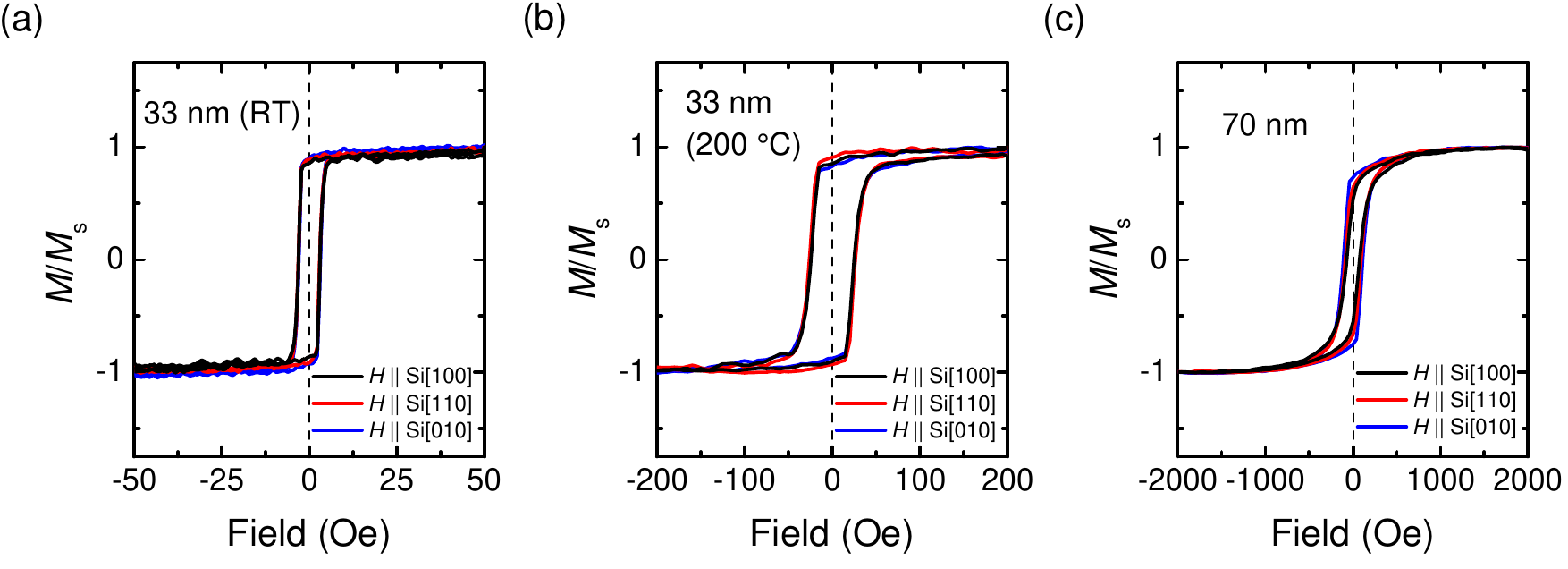}
  \caption{Vibrating sample magnetometry of (a) 33~nm (room temperature deposition), (b) 33~nm (200~$^\circ$C deposition), and (c) 70~nm films for $H \parallel \textrm{Si}[100]$ (black), $H \parallel \textrm{Si}[110]$ (red), and $H \parallel \textrm{Si}[010]$ (blue).}\label{fig:VSM}
\end{figure}

Vibrating sample magnetometry (VSM) data for the 33~nm (room temperature and 200~$^\circ$C depositions) and 70~nm films are shown in Fig.\ \ref{fig:VSM}. The magnetic field was applied in 3 different directions, with no discernible difference in the hysteresis loops. We conclude that there is no in-plane magnetocrystalline anisotropy over macroscopic lengthscales, which is consistent with the FMR measurements.
\section{Longitudinal resistivity}
\begin{figure}
  \centering
  \includegraphics{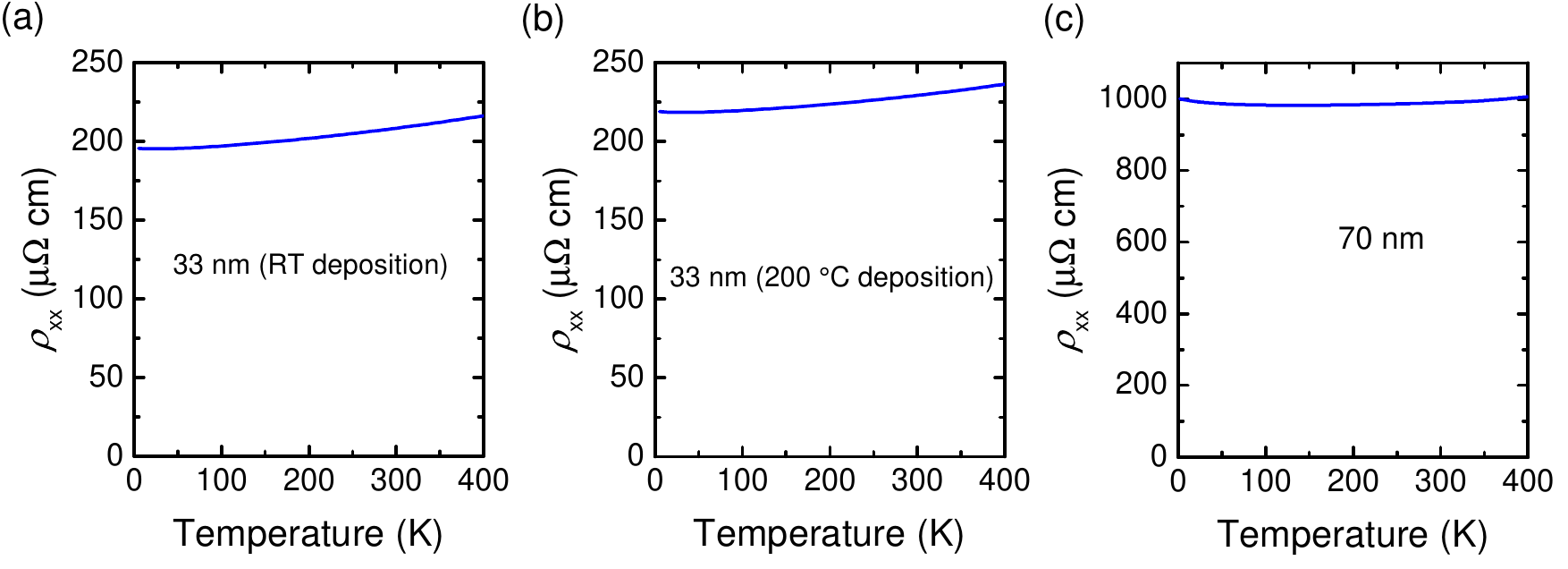}
  \caption{Longitudinal resistivity $\rho_{xx}$ as a function of temperature for the (a) 33 nm (room temperature deposition), (b) 33 nm (200~$^\circ$C deposition), and (c) 70~nm films.}\label{fig:rhoxx}
\end{figure}

Longitudinal resistivity $\rho_{xx}$ was measured as a function of temperature for the 33~nm (room temperature and 200~$^\circ$C depositions) and 70~nm films (Fig.\ \ref{fig:rhoxx}) by patterning Hall bars and performing 4-wire resistance measurements.
%
%
%